\documentclass[a4paper, table]{article}

\usepackage{itgspeech2023}    %
\usepackage{times}            %
\usepackage[english]{babel}   %
\usepackage[ansinew]{inputenc}%
\usepackage[T1]{fontenc}      %
\usepackage[sort&compress,numbers]{natbib}	%
\usepackage{amsmath,amssymb}
\usepackage{graphicx}
\usepackage[colorlinks=false,pdfborder={0 0 0}]{hyperref}
\usepackage{units}
\usepackage[nolist, nohyperlinks]{acronym}
\usepackage{makecell}
\usepackage{highlight}
\usepackage{booktabs}
\usepackage{flushend}
\usepackage{microtype}

\renewcommand{\opacity}{70}

\title{On the Behavior of Intrusive and Non-intrusive Speech Enhancement Metrics in Predictive and Generative Settings}

\author{Danilo de Oliveira, Julius Richter, Jean-Marie Lemercier, Tal Peer, Timo Gerkmann}

\address{Universit\"at Hamburg, Hamburg, Germany\\
  Email: \texttt{danilo.oliveira@uni-hamburg.de}}%

\begin{document}

\begin{acronym}
\acro{sgm}[SGM]{score-based generative model}
\acro{snr}[SNR]{signal-to-noise ratio}
\acro{gan}[GAN]{generative adversarial network}
\acro{vae}[VAE]{variational autoencoder}
\acro{ddpm}[DDPM]{denoising diffusion probabilistic model}
\acro{stft}[STFT]{short-time Fourier transform}
\acro{istft}[iSTFT]{inverse short-time Fourier transform}
\acro{sde}[SDE]{stochastic differential equation}
\acro{ode}[ODE]{ordinary differential equation}
\acro{ou}[OU]{Ornstein-Uhlenbeck}
\acro{ve}[VE]{Variance Exploding}
\acro{dnn}[DNN]{deep neural network}
\acro{pesq}[PESQ]{Perceptual Evaluation of Speech Quality}
\acro{se}[SE]{speech enhancement}
\acro{tf}[T-F]{time-frequency}
\acro{elbo}[ELBO]{evidence lower bound}
\acro{WPE}{weighted prediction error}
\acro{PSD}{power spectral density}
\acro{RIR}{room impulse response}
\acro{LSTM}{long short-term memory}
\acro{POLQA}{Perceptual Objectve Listening Quality Analysis}
\acro{SDR}{signal-to-distortion ratio}
\acro{ESTOI}{Extended Short-Term Objective Intelligibility}
\acro{ELR}{early-to-late reverberation ratio}
\acro{TCN}{temporal convolutional network}
\acro{DRR}{direct-to-reverberant ratio}
\acro{nfe}[NFE]{number of function evaluations}
\acro{rtf}[RTF]{real-time factor}
\acro{mos}[MOS]{mean opinion scores}
\acro{asr}[ASR]{automatic speech recognition}
\acro{wer}[WER]{word error rate}
\acro{avse}[AVSE]{audio-visual speech enhancement}
\acro{ssl}[SSL]{self-supervised learning}
\acro{dl}[DL]{deep learning}
\acro{roi}[ROI]{region-of-interest}
\acro{ncsn}[NCSN++]{Noise Conditional Score Network}
\end{acronym}

\maketitle

\begin{abstract}

Since its inception, the field of deep speech enhancement has been dominated by predictive (discriminative) ap\-proach\-es, such as spectral mapping or masking. Recently, however, novel generative approaches have been applied to speech enhancement, attaining good denoising performance with high subjective quality scores. At the same time, advances in deep learning also allowed for the creation of neural network-based metrics, which have desirable traits such as being able to work without a reference (non-intrusively). Since generatively enhanced speech tends to exhibit radically different residual distortions, its evaluation using instrumental speech metrics may behave differently compared to predictively enhanced speech. In this paper, we evaluate the performance of the same speech enhancement backbone trained under predictive and generative paradigms on a variety of metrics and show that intrusive and non-intrusive measures correlate differently for each paradigm. This analysis motivates the search for metrics that can together paint a complete and unbiased picture of speech enhancement performance, irrespective of the model's training process.

\end{abstract}

\section{Introduction}

Advances in \acp{dnn} in the past de\-cades have greatly benefited tasks in a variety of fields, most notably computer vision \cite{krizhevsky2012alexnet, he2016resnet}, natural language processing \cite{devlin2019bert} and speech processing/recognition \cite{hinton2012deep}. Included in the latter is the task of speech enhancement, which experienced a shift from traditional statistical based methods \cite{ephraim1984speech, breithaupt2007cepstral} to models that learn audio denoising from data itself. This has led to great improvements in speech quality and intelligibility (correct identification of phonemes/words), in particular in the case of non-stationary noise. With a few exceptions such as variational autoencoders \cite{fang2021variational, carbajal2021guided}, \ac{dnn}-based speech enhancement has generally been dominated by predictive (also known as discriminative) models, which given a noisy speech input, map it to a single clean speech estimate \cite{lemercier2023analysing}. Popular techniques in the domain are spectral mapping \cite{tan2020learning} or masking \cite{hu2020dccrn, deoliveira2022efficient}, as well as time domain approaches \cite{defossez2020realtime}.

Recently, the development of generative approaches for speech enhancement has made significant progress, leading to competitive performance with good generalization capability compared to predictive models \cite{richter2023speech}. Instead of learning a direct mapping from noisy speech to the clean target, they learn to model the underlying statistical distribution of clean speech, typically involving some sort of stochasticity in the form of random variables. Given this different learning paradigm, generative speech enhancement models show distinct behavior and introduce other types of distortions. Typically the enhanced speech will resemble the characteristics of the clean speech training data. However, since conditional generative models take the noisy speech as an auxiliary input, they tend to hallucinate for signals with low \acp{snr}, which can lead to speech-like sounds with poor articulation and no linguistic meaning \cite{richter2023speech}.

In order to compare the myriad of models that are created by the research community and to allow for faster development of new, improved methods, instrumental evaluation metrics are indispensable. Although it is desirable to obtain a model whose output is pleasant to humans, subjective evaluations are impractical in most settings, as conducting extensive listening experiments can be expensive and time-consuming. Therefore, the use of instrumental metrics is preferred when developing speech enhancement systems. These metrics are designed to serve as proxy for subjective measures such as intelligibility or quality of the enhanced audio. Subjective quality is a more challenging measure to quantify \cite{loizou2013speech}, and is commonly evaluated by means of Perceptual Evaluation of Speech Quality (PESQ) \cite{rix2001pesq} or Perceptual Objective Listening Quality Analysis (POLQA) \cite{beerends2013perceptual}. Both these metrics are \emph{intrusive}, i.e. in addition to the corrupted or enhanced signal being evaluated, they require a reference clean speech signal.

Neural networks, due to their versatility and capability of modeling complex patterns, are also good candidates to tackle the task of modeling human perception. Consequently, their use has been extended beyond the suppression of noise and into the evaluation of noise suppression algorithms. A major advantage of these methods is that they are often trained to work in a \emph{non-intrusive} manner, i.e. without a reference signal. This extends their applicability beyond curated speech corpora and allows for the evaluation of real-world data. Models like DNSMOS \cite{reddy2022dnsmosp835} and WV-MOS \cite{andreev2023hifipp}, given an audio sample, estimate a scalar value for the speech quality or naturalness in the mean opinion score (MOS) scale.

Due to the different characteristics of the estimated speech signals and the different nature of residual artifacts, instrumental metrics can be expected to behave differently when evaluating predictive or generative models. The purpose of this paper is to analyze the behavior of different intrusive and non-intrusive speech metrics on the output of a predictive and a generative model that share
the same network architecture. We offer insight into how each metric takes different distortions into account and establish the need for new evaluation metrics that would capture the characteristics of both predictive and generative approaches, and allow for fair comparison between them.

\section{Comparing Predictive and Generative Speech Enhancement}

While predictive models have shown impressive performance owing to advances in network architectures, specialized objective functions, and ever-larger models and datasets, they still often suffer from distortions of the output signal and generalization issues \cite{wang2019bridging, richter2023speech}. These distortions have certain characteristics that can be detected with instrumental metrics by measuring some numerical distance between clean and enhanced speech signals \cite{loizou2013speech}. For instance, measuring a distance to assess speech quality typically includes an auditory model which is designed to resemble human perception. Non-intrusive metrics are also capable of identifying certain distortions or assessing speech quality as long as the underlying models are trained with the respective data. In other words, non-intrusive metrics learn to look for patterns that resemble the types of distortions that were presented at training. Intrusive metrics, on the other hand, are always biased toward the reference signal and output the maximum score if the estimate corresponds exactly to the clean speech target.
 
In contrast, generative models implicitly or explicitly learn the statistical distribution of clean speech, enabling the generation of multiple possible clean speech estimates. These estimates adhere to the learned distribution of clean speech, and should ideally resemble factors such as speaker characteristics, prosody, and semantic content inferred from the conditional noisy input. However, in adverse noise conditions where the noise component dominates over the clean speech, conditional generative models tend to produce speech-shaped noise or introduce some hallucinations \cite{lemercier2023analysing}. This can lead to speech inpainting in parts where there is no speech originally, or to the introduction of phonetic confusions. In this situation, metrics that prioritize exact signal-level reconstruction would typically experience a decline in performance, although human listeners do not necessarily penalize these effects as harshly \cite{richter2023speech}. However, non-intrusive metrics do not have access to the reference signal and may not be able to detect these artifacts since they are speech-like in nature.

To systematically investigate how different metrics behave in this context, we analyze the speech enhancement performance of a predictive model and a generative model. Both models share the same neural network architecture but follow different learning and inference paradigms which we will describe below. For the analysis, we measure various intrusive and non-intrusive metrics, as described in Sec. \ref{subsec:metrics}.

\subsection{Speech Enhancement Models}

\subsubsection{Generative Model}

We use a score-based generative model \cite{richter2023speech, lemercier2023analysing} based on a continuous-time diffusion process \cite{song2021scorebased}. During training, a forward diffusion model turns the clean data into a tractable noise distribution by progressively adding Gaussian and environmental noise. The so-called \textit{score} function i.e. the gradient of the logarithmic data distribution, is learned by the neural network called \textit{score model} using a denoising score matching objective \cite{vincent2011connection}. At inference time, the score function estimate obtained via the neural network is used to solve the reverse diffusion process \cite{anderson1982reverse}. This reverse diffusion process iteratively removes the environmental noise and some previously added Gaussian noise, and finally generates a clean speech estimate.

\subsubsection{Predictive Model}

The predictive model uses the same network architecture of the generative model but is simply trained to perform a mapping from the noisy complex spectrogram to the clean complex spectrogram. The objective function used to tune the model parameters is the classical mean squared error, applied in the complex spectrogram domain.

\subsection{Evaluation Metrics}\label{subsec:metrics}

\subsubsection{Intrusive}

\begin{itemize}
    \item \textbf{ESTOI}: Extended Short-Time Objective Intelligibility \cite{jensen2016algorithm} is a metric for speech intelligibility. It builds up on its predecessor STOI \cite{taal2011algorithm}, which works based on the correlation between temporal envelopes of clean and processed speech in short-time segments. ESTOI models intelligibility through a ranking of features from decomposed energy-normalized spectrograms and works well also for highly modulated noise sources. The metric outputs a score in the range $[0,1]$.
    
    \item \textbf{POLQA}: Perceptual Objective Listening Quality Analysis \cite{beerends2013perceptual} is a metric for speech quality. It is the successor to PESQ \cite{rix2001pesq}, a popular metric designed for the assessment of telephone networks and codecs. POLQA is additionally suited for the evaluation of speech enhancement systems making use of non-linear processing. POLQA works by comparing the reference with the aligned processed signal using an internal representation from a perceptual model, and outputting a value in the MOS scale, from 1 to 5. We use the wideband version of POLQA.
    
    \item \textbf{SI-SDR}: Scale-invariant signal-to-distortion ratio \cite{roux2019sdr} measures distortions in the signal, with the advantage of accounting for scaling discrepancies. Its calculation is simple and it has been used extensively as a training-loss. However, SNR-related metrics such as SI-SDR have a limited ability to predict subjective human perception \cite{loizou2013speech}.
    
    \item \textbf{WAcc}: Similarly to \cite{dubey2022dnschallenge}, word accuracy is defined here as $\mathrm{WAcc} = \min(1 - \mathrm{WER}, 0)$, where WER is the word error rate, a metric commonly used in the evaluation of automatic speech recognition (ASR) systems. WER is the ratio between the number of inserted, deleted and replaced words over the total number of words in the reference utterance. In order to evaluate WAcc, an ASR model is needed to generate transcriptions from the enhanced speech samples. We use the \texttt{base-en} version of Quartznet \cite{kriman2020quartznet} as the ASR system.

\end{itemize}

\subsubsection{Non-intrusive}

\begin{itemize}
    \item \textbf{DNSMOS P.835}: DNSMOS P.835 \cite{reddy2022dnsmosp835} uses a convolutional neural network to predict MOS scores given by the ITU-T P. 835 standard. In this standard, participants are asked to rate on a MOS scale the signal quality (DNSMOS SIG), reduction of background noise (DNSMOS BAK) and overall audio quality (DNSMOS OVRL) of samples given in a listening test. The listening test used DNS Challenge data \cite{dubey2022dnschallenge} and presented many subjects with speech stemming from the clean and noisy subsets as well as outputs of many denoising methods.
    
    \item \textbf{WV-MOS}: WV-MOS uses a pre-trained wav2vec2.0 self-supervised model \cite{baevski2020wav2vec2} and two fine-tuned fully connected layers to predict assigned MOS scores in a voice conversion task \cite{andreev2023hifipp}. It has been shown by the authors to be relevant for speech denoising as well as bandwidth extension.
    
\end{itemize}
\begingroup

\setlength{\tabcolsep}{4pt} %
\renewcommand{\arraystretch}{1.4} %
\begin{table*}[t]
	\centering
 \scalebox{0.95}{
	\resizebox{\textwidth}{!}{\begin{tabular}{l c c c c c c c c}
  \toprule
    & \multicolumn{4}{c}{Intrusive} & \multicolumn{4}{c}{Non-intrusive} \\ \cmidrule(lr){2-5} \cmidrule(lr){6-9}
                & SI-SDR & ESTOI & WAcc & POLQA & \makecell{DNSMOS\\SIG} & \makecell{DNSMOS\\BAK} & \makecell{DNSMOS\\OVRL} & WV-MOS \\
		\cmidrule(lr){2-9}
		Noisy   & $3.8\pm5.8$ & $0.56\pm0.17$ & $0.66\pm0.34$ & $1.99\pm0.66$ & $2.67\pm0.83$ & $2.32\pm0.84$ & $2.02\pm0.61$ & $0.50\pm2.19$ \\
		NCSN++M & $\mathbf{15.1\pm5.0}$ & $\mathbf{0.81\pm0.13}$ & $\mathbf{0.75\pm0.28}$ & $\mathbf{3.08\pm0.81}$ & $3.24\pm0.29$ & $\mathbf{3.95\pm0.20}$ & $\mathbf{2.97\pm0.31}$ & $3.05\pm0.61$ \\
        SGMSE+M & $12.9\pm5.5$ & $0.79\pm0.15$ & $0.68\pm0.32$ & $2.99\pm0.79$ & $\mathbf{3.42\pm0.15}$ & $3.65\pm0.46$ & $\mathbf{2.97\pm0.31}$ & $\mathbf{3.46\pm0.67}$ \\
  \bottomrule
	\end{tabular}}}
	\caption{Speech enhancement evaluated on TIMIT+CHiME. NCSN++M corresponds to predictive objective modeling, while SGMSE+M represents the generative modeling approach.}
	\label{tab:absolute_results}%
\end{table*}
\endgroup

\begin{figure*}[t]
	\centerline{\includegraphics[width=\textwidth]{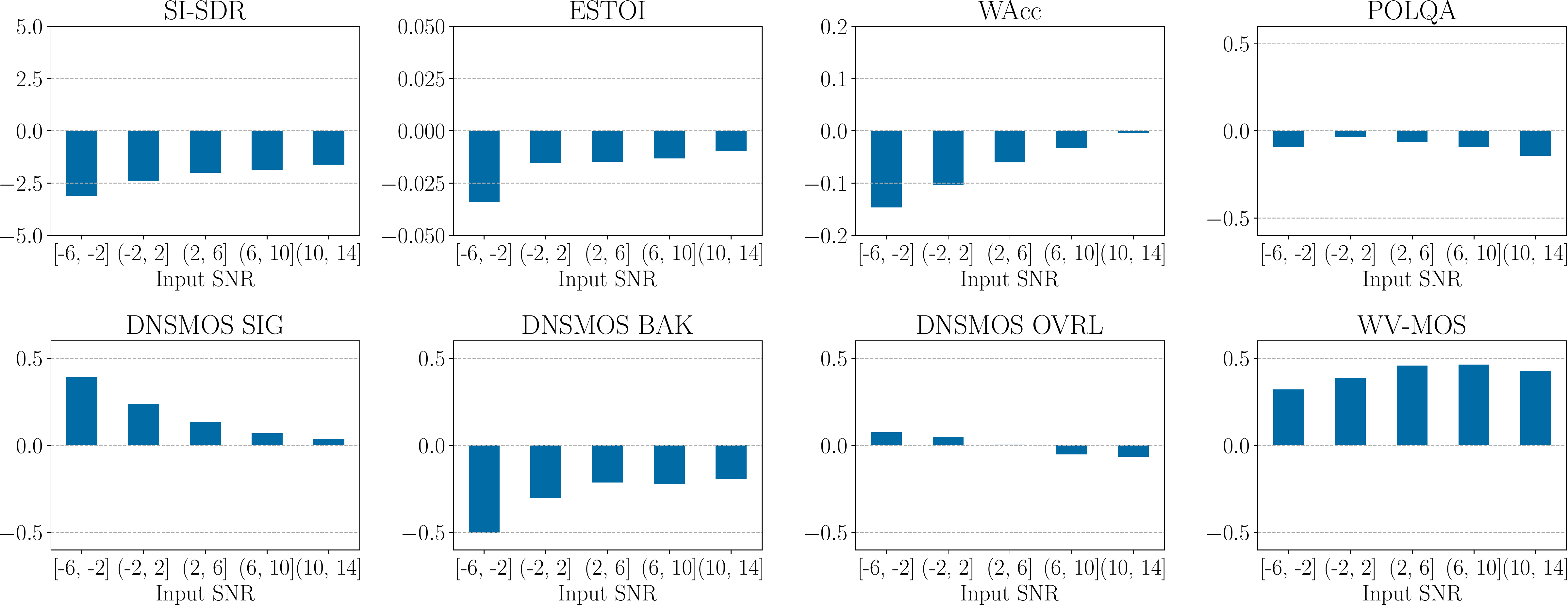}}
	\caption{Differences between evaluation metric values for generative and predictive models, grouped by input SNR. Positive/negative values indicate better performance of the generative/predictive model, respectively.}
	\label{fig:diff}%
\end{figure*} 

\section{Implementation Details}

\subsection{Data}\vspace{-2px}

We generate the TIMIT+CHiME3 dataset using TIMIT \cite{garofolo1993timit} as the clean speech corpus and noise samples from the CHiME3 challenge data \cite{barker2015chime} containing four types of noise: caf\'e, street, pedestrian and bus noise. Utterances are randomly picked and mixed additively with SNR uniformly distributed between -6 and 14dB. In total, 3.1, 0.6 and 1.1 hours of clean and noisy speech sampled at 16kHz are generated for training, validation, and test respectively.

Noisy and clean utterances are transformed using a STFT with a window size of 510, a hop length of 128 and a square-root Hann window, at a sampling rate of $16$kHz, as in \cite{richter2023speech}. A square-root magnitude warping is used to reduce the dynamic range of the spectrograms.
During training, sequences of 256 \ac{stft} frames ($\sim$2s) are extracted from the full-length utterances with random offsets and normalized in the time-domain by the maximum absolute value of the noisy utterance. 

\subsection{Neural Network Architecture}\vspace{-2px}

For both the predictive and the generative model we use a lighter configuration of the \ac{ncsn} \cite{song2021scorebased}, which was proposed in \cite{lemercier2023analysing} and denoted as NCSN++M, with roughly 27.8M parameters.
In the predictive approach, the noisy speech spectrogram's real and imaginary parts are stacked and passed as sole input to the network, and no noise-conditioning is used. In the generative model, denoted as \emph{SGMSE+M}, score estimation during reverse diffusion is carried out by stacking the real and imaginary parts of the noisy speech spectrogram and of the current diffusion estimate, feeding them to the score network along with the current noise level as a conditioner.

\subsection{Hyperparameters and Training Details}\vspace{-2px}

As in \cite{lemercier2023analysing}, we train NCSN++M and SGMSE+M  using the Adam optimizer \cite{kingma2015adam} with a learning rate of $5\cdot10^{-4}$ and an effective batch size of 16. We track an exponential moving average of the DNN weights with a decay of 0.999. Early stopping based on the validation loss is used with a patience of 10 epochs.
The configuration of the diffusion process and the reverse sampler is identical to \cite{richter2023speech}.

\begin{table*}[t]
	\centering
 \scalebox{0.95}{
	\begin{tabular}[0.9\textwidth]{l *{8}{c}}
        \multicolumn{9}{c}{Predictive (NCSN++M)}\\
  \toprule
                    & Input SNR & SI-SDR & ESTOI & POLQA & WAcc & WV-MOS & \makecell{DNSMOS\\SIG} & \makecell{DNSMOS\\BAK} \\
        \cmidrule{2-9}
        SI-SDR      & \gradient{0.822} & & & & & & & \\
        ESTOI       & \gradient{0.770} & \gradient{0.870} &	& & & & \\
        POLQA       & \gradient{0.779} & \gradient{0.929} & \gradient{0.882} & & & & & \\
        WAcc        & \gradient{0.566} & \gradient{0.655} & \gradient{0.741} & \gradient{0.685} & & & & \\
        WV-MOS      & \gradient{0.629} & \gradient{0.759} & \gradient{0.785} & \gradient{0.801} & \gradient{0.613} & & & \\
        DNSMOS SIG  & \gradient{0.630} & \gradient{0.704} & \gradient{0.781} & \gradient{0.717} & \gradient{0.647} & \gradient{0.647} & & \\
        DNSMOS BAK  & \gradient{0.447} & \gradient{0.494} & \gradient{0.495} & \gradient{0.501} & \gradient{0.402} & \gradient{0.411} & \gradient{0.625} \\
        DNSMOS OVRL & \gradient{0.632} & \gradient{0.703} & \gradient{0.764} & \gradient{0.714} & \gradient{0.631} & \gradient{0.631} & \gradient{0.971} &	\gradient{0.788} \\
        
  \bottomrule
	\end{tabular}}
 \newline
\vspace{5pt}
\newline
    \scalebox{0.95}{
  \begin{tabular}[0.9\textwidth]{l *{8}{c}}
        \multicolumn{9}{c}{Generative (SGMSE+M)}\\
  \toprule
                    & Input SNR & SI-SDR & ESTOI & POLQA & WAcc & WV-MOS & \makecell{DNSMOS\\SIG} & \makecell{DNSMOS\\BAK} \\
        \cmidrule{2-9}
        SI-SDR      & \gradient{0.829} & & & & & & & \\
        ESTOI       & \gradient{0.751} & \gradient{0.875} & & & & & \\
        POLQA       & \gradient{0.775} & \gradient{0.908} & \gradient{0.899} & & & & & \\
        WAcc        & \gradient{0.636} & \gradient{0.727} & \gradient{0.803} & \gradient{0.784} & & & & \\
        WV-MOS      & \gradient{0.638} & \gradient{0.760} & \gradient{0.804} & \gradient{0.769} & \gradient{0.692} & & & \\
        DNSMOS SIG  & \gradient{0.328} & \gradient{0.388} & \gradient{0.423} & \gradient{0.400} & \gradient{0.349} & \gradient{0.342} & & \\
        DNSMOS BAK  & \gradient{0.429} & \gradient{0.570} & \gradient{0.525} & \gradient{0.525} & \gradient{0.417} & \gradient{0.475} & \gradient{0.501} & \\
        DNSMOS OVRL & \gradient{0.461} & \gradient{0.589} & \gradient{0.569} & \gradient{0.561} & \gradient{0.462} & \gradient{0.499} & \gradient{0.749} & \gradient{0.940} \\
        
  \bottomrule
	\end{tabular}}
	\caption{Pearson correlation between evaluation metrics on the denoising results for TIMIT+CHiME. 0 corresponds to no linear correlation, while 1 would represent maximal linear correlation.}
	\label{tab:correlations}%
\end{table*}

\section{Results and Discussion}\vspace{-3px}

We evaluate speech utterances processed by NCSN++M (predictive) and by SGMSE+M (generative) according to the metrics listed in Sec.~\ref{subsec:metrics}. Table~\ref{tab:absolute_results} shows the mean and standard deviation values for each model. We can note that intrusive metric scores are better in the predictive approach, while non-intrusive ones mostly favor the generative model. Although DNSMOS SIG scores are higher for SGMSE+M, DNSMOS BAK is lower than for NCSN++M. This suggests that the generative model produces less degraded, more natural speech, but the predictive model performs more suppression of background noise. With both aspects tending to balance each other, the models perform fairly equally on DNSMOS OVRL.

Aiming at a better understanding of in the models' differences in handling smaller and larger amounts of noise, we look at the results grouped by SNR of the noisy mixtures. Fig.~\ref{fig:diff} features the differences in performance between the generative and the predictive approach across input SNRs. Positive values are cases in which the generative model fares better, while negative values indicate better performance of the predictive model. We can see that the gap in intrusive metrics between the predictive and the generative model increases as the input SNR gets lower. This could be explained by the hallucinations introduced by generative models, penalized more severely by intrusive metrics that expect a signal-level match to the reference.

In order to visualize how metrics correlate with one another, we compute the Pearson correlation coefficient (PCC) between each of them, pair-wise. We also compute the PCC of each metric with input SNR. These coefficients are presented in Table~\ref{tab:correlations}, where the value of the cell represents the correlation between the corresponding row and column. SI-SDR, POLQA, and ESTOI present strong correlations with each other in both predictive and generative settings. WAcc correlation with intrusive metrics and WV-MOS is higher in the generative setting, suggesting that the distortions introduced by generative models that are penalized by intrusive measures also have a larger impact on what is perceived by the ASR model. In the generative paradigm, DNSMOS SIG and OVRL metrics have significantly lower correlations with other metrics and input SNR. This indicates that the estimates generated by SGMSE+M have the characteristics of clean speech, even if they doesn't necessarily match the references and don't have coherent language content. 

Though it is also a non-intrusive metric, WV-MOS does not follow the correlation behavior of DNSMOS under the generative paradigm. We hypothesize that this is due to the semantic information implicitly modeled in the inner representations of the pre-trained self-supervised model \cite{pasad2021layerwise}. Hallucinations generated by generative models could sound as high-quality speech to DNSMOS, while the language modeling capabilities of wav2vec2.0 would spot and penalize semantic inconsistencies in the generated audio.

These results suggest that a combination of non-in\-tru\-sive metrics with metrics that evaluate the linguistic content preservation is beneficial. In \cite{dubey2022dnschallenge}, denoising performance is ranked based on the average between DNSMOS OVRL and WAcc (both scaled to the same range). 
However, reliable evaluation of WAcc requires transcription labels, often not available. Additionally, there may be metrics more suitable to evaluate preservation of speech content in the context of distortions introduced by the different paradigms. More research in finding existing metrics or creating new ones with that intent will greatly benefit the continuous development of better speech enhancement systems. 

\section{Conclusions}\vspace{-3px}

In this paper, we studied the behavior of the same speech enhancement \ac{dnn} trained under two different paradigms: predictive and generative. We evaluated the models using a variety of instrumental metrics and analyzed how the different metrics treat the different model paradigms. We found that the generative model is often favored by non-intrusive metrics of speech quality/natural\-ness, while being penalized to a greater extent by intrusive metrics, in particular at low input SNRs. 
Based on these results, we argue that there is a need for further studies addressing the preservation of semantic and phonetic content in speech enhancement, as well as for novel metrics that can detect both speech distortions and generative artifacts. Such metrics would allow proper evaluation of generative approaches by offering a better correlation with human listening experience.

\section{Acknowledgment}\vspace{-3px}

This work has been funded by the German Research Foundation (DFG) in the transregio project Crossmodal Learning (TRR 169). We would like to thank J. Berger and Rohde\&Schwarz SwissQual AG for their support with POLQA.

\clearpage
\small
\atColsBreak{\vskip1pt}
\bibliographystyle{ieeetr}
\bibliography{references}

\begin{thebibliography}{10}

\bibitem{krizhevsky2012alexnet}
A.~Krizhevsky, I.~Sutskever, and G.~E. Hinton, ``Imagenet classification with
  deep convolutional neural networks,'' in {\em Advances in Neural Inf. Proc.
  Systems (NeurIPS)}, vol.~25, 2012.

\bibitem{he2016resnet}
K.~He, X.~Zhang, S.~Ren, and J.~Sun, ``Deep residual learning for image
  recognition,'' in {\em IEEE/CVF Conf. on Computer Vision and Pattern
  Recognition (CVPR)}, June 2016.

\bibitem{devlin2019bert}
J.~Devlin, M.-W. Chang, K.~Lee, and K.~Toutanova, ``{BERT}: {Pre}-training of
  {Deep} {Bidirectional} {Transformers} for {Language} {Understanding},'' in
  {\em Proceedings of the 2019 {Conference} of the {North} {American} {Chapter}
  of the {Association} for {Computational} {Linguistics}: {Human} {Language}
  {Technologies} ({NAACL-HLT})}, vol.~1, pp.~4171--4186, 2019.

\bibitem{hinton2012deep}
G.~Hinton, L.~Deng, D.~Yu, G.~E. Dahl, A.-r. Mohamed, N.~Jaitly, A.~Senior,
  V.~Vanhoucke, P.~Nguyen, T.~N. Sainath, and B.~Kingsbury, ``Deep neural
  networks for acoustic modeling in speech recognition: The shared views of
  four research groups,'' {\em IEEE Signal Processing Magazine}, vol.~29,
  no.~6, pp.~82--97, 2012.

\bibitem{ephraim1984speech}
Y.~Ephraim and D.~Malah, ``{Speech Enhancement Using a Minimum Mean-Square
  Error Short-Time Spectral Amplitude Estimator},'' {\em IEEE Trans. on Audio,
  Speech, and Lang. Process. (TASLP)}, vol.~ASSP-32, pp.~1109--1121, Dec. 1984.

\bibitem{breithaupt2007cepstral}
C.~Breithaupt, T.~Gerkmann, and R.~Martin, ``Cepstral smoothing of spectral
  filter gains for speech enhancement without musical noise,'' {\em IEEE Signal
  Process. Lett. (SPL)}, vol.~14, no.~12, pp.~1036--1039, 2007.

\bibitem{fang2021variational}
H.~Fang, G.~Carbajal, S.~Wermter, and T.~Gerkmann, ``Variational autoencoder
  for speech enhancement with a noise-aware encoder,'' in {\em IEEE Int. Conf.
  on Acoustics, Speech and Signal Process. (ICASSP)}, pp.~676--680, 2021.

\bibitem{carbajal2021guided}
G.~Carbajal, J.~Richter, and T.~Gerkmann, ``Guided variational autoencoder for
  speech enhancement with a supervised classifier,'' in {\em IEEE Int. Conf. on
  Acoustics, Speech and Signal Process. (ICASSP)}, pp.~681--685, 2021.

\bibitem{lemercier2023analysing}
J.-M. Lemercier, J.~Richter, S.~Welker, and T.~Gerkmann, ``Analysing
  diffusion-based generative approaches versus discriminative approaches for
  speech restoration,'' in {\em IEEE Int. Conf. on Acoustics, Speech and Signal
  Process. (ICASSP)}, pp.~1--5, 2023.

\bibitem{tan2020learning}
K.~Tan and D.~Wang, ``Learning complex spectral mapping with gated
  convolutional recurrent networks for monaural speech enhancement,'' {\em IEEE
  Trans. on Audio, Speech, and Lang. Process. (TASLP)}, vol.~28, pp.~380--390,
  2020.

\bibitem{hu2020dccrn}
Y.~Hu, Y.~Liu, S.~Lv, M.~Xing, S.~Zhang, Y.~Fu, J.~Wu, B.~Zhang, and L.~Xie,
  ``{DCCRN: Deep Complex Convolution Recurrent Network for Phase-Aware Speech
  Enhancement},'' in {\em Interspeech}, pp.~2472--2476, 2020.

\bibitem{deoliveira2022efficient}
D.~{de Oliveira}, T.~Peer, and T.~Gerkmann, ``{Efficient Transformer-based
  Speech Enhancement Using Long Frames and STFT Magnitudes},'' in {\em
  Interspeech}, pp.~2948--2952, 2022.

\bibitem{defossez2020realtime}
A.~D\'efossez, G.~Synnaeve, and Y.~Adi, ``{Real Time Speech Enhancement in the
  Waveform Domain},'' in {\em Interspeech}, pp.~3291--3295, 2020.

\bibitem{richter2023speech}
J.~Richter, S.~Welker, J.-M. Lemercier, B.~Lay, and T.~Gerkmann, ``Speech
  enhancement and dereverberation with diffusion-based generative models,''
  {\em IEEE Trans. on Audio, Speech, and Lang. Process. (TASLP)}, 2023.

\bibitem{loizou2013speech}
P.~Loizou, {\em Speech Enhancement: Theory and Practice, Second Edition}.
\newblock CRC Press, 2013.

\bibitem{rix2001pesq}
A.~Rix, J.~Beerends, M.~Hollier, and A.~Hekstra, ``Perceptual evaluation of
  speech quality (pesq)-a new method for speech quality assessment of telephone
  networks and codecs,'' in {\em IEEE Int. Conf. on Acoustics, Speech and
  Signal Process. (ICASSP)}, vol.~2, pp.~749--752 vol.2, 2001.

\bibitem{beerends2013perceptual}
J.~G. Beerends, C.~Schmidmer, J.~Berger, M.~Obermann, R.~Ullmann, J.~Pomy, and
  M.~Keyhl, ``Perceptual objective listening quality assessment (polqa), the
  third generation itu-t standard for end-to-end speech quality measurement
  part i - temporal alignment,'' {\em Journal of the Audio Engineering Society
  (AES)}, vol.~61, no.~6, pp.~366--384, 2013.

\bibitem{reddy2022dnsmosp835}
C.~K.~A. Reddy, V.~Gopal, and R.~Cutler, ``Dnsmos p.835: A non-intrusive
  perceptual objective speech quality metric to evaluate noise suppressors,''
  in {\em IEEE Int. Conf. on Acoustics, Speech and Signal Process. (ICASSP)},
  pp.~886--890, 2022.

\bibitem{andreev2023hifipp}
P.~Andreev, A.~Alanov, O.~Ivanov, and D.~Vetrov, ``Hifi++: A unified framework
  for bandwidth extension and speech enhancement,'' in {\em IEEE Int. Conf. on
  Acoustics, Speech and Signal Process. (ICASSP)}, pp.~1--5, 2023.

\bibitem{wang2019bridging}
P.~Wang, K.~Tan, and D.~L. Wang, ``Bridging the gap between monaural speech
  enhancement and recognition with distortion-independent acoustic modeling,''
  {\em IEEE Trans. on Audio, Speech, and Lang. Process. (TASLP)}, vol.~28,
  pp.~39--48, 2020.

\bibitem{song2021scorebased}
Y.~Song, J.~Sohl-Dickstein, D.~P. Kingma, A.~Kumar, S.~Ermon, and B.~Poole,
  ``Score-based generative modeling through stochastic differential
  equations,'' in {\em Int. Conf. on Learning Representations (ICLR)}, 2021.

\bibitem{vincent2011connection}
P.~Vincent, ``A connection between score matching and denoising autoencoders,''
  {\em Neural Computation}, vol.~23, no.~7, pp.~1661--1674, 2011.

\bibitem{anderson1982reverse}
B.~D. Anderson, ``Reverse-time diffusion equation models,'' {\em Stochastic
  Processes and their Applications}, vol.~12, no.~3, pp.~313--326, 1982.

\bibitem{jensen2016algorithm}
J.~Jensen and C.~H. Taal, ``An algorithm for predicting the intelligibility of
  speech masked by modulated noise maskers,'' {\em IEEE Trans. on Audio,
  Speech, and Lang. Process. (TASLP)}, vol.~24, no.~11, pp.~2009--2022, 2016.

\bibitem{taal2011algorithm}
C.~H. Taal, R.~C. Hendriks, R.~Heusdens, and J.~Jensen, ``An algorithm for
  intelligibility prediction of time-frequency weighted noisy speech,'' {\em
  IEEE Trans. on Audio, Speech, and Lang. Process. (TASLP)}, vol.~19, no.~7,
  pp.~2125--2136, 2011.

\bibitem{roux2019sdr}
J.~L. Roux, S.~Wisdom, H.~Erdogan, and J.~R. Hershey, ``Sdr - half-baked or
  well done?,'' in {\em IEEE Int. Conf. on Acoustics, Speech and Signal
  Process. (ICASSP)}, pp.~626--630, 2019.

\bibitem{dubey2022dnschallenge}
H.~Dubey, V.~Gopal, R.~Cutler, A.~Aazami, S.~Matusevych, S.~Braun, S.~E.
  Eskimez, M.~Thakker, T.~Yoshioka, H.~Gamper, and R.~Aichner, ``Icassp 2022
  deep noise suppression challenge,'' in {\em IEEE Int. Conf. on Acoustics,
  Speech and Signal Process. (ICASSP)}, pp.~9271--9275, 2022.

\bibitem{kriman2020quartznet}
S.~Kriman, S.~Beliaev, B.~Ginsburg, J.~Huang, O.~Kuchaiev, V.~Lavrukhin,
  R.~Leary, J.~Li, and Y.~Zhang, ``Quartznet: Deep automatic speech recognition
  with 1d time-channel separable convolutions,'' in {\em IEEE Int. Conf. on
  Acoustics, Speech and Signal Process. (ICASSP)}, pp.~6124--6128, 2020.

\bibitem{baevski2020wav2vec2}
A.~Baevski, Y.~Zhou, A.~Mohamed, and M.~Auli, ``wav2vec 2.0: A framework for
  self-supervised learning of speech representations,'' in {\em Advances in
  Neural Inf. Proc. Systems (NeurIPS)}, vol.~33, pp.~12449--12460, 2020.

\bibitem{garofolo1993timit}
J.~S. {Garofolo}, L.~F. {Lamel}, W.~M. {Fisher}, J.~G. {Fiscus}, and D.~S.
  {Pallett}, ``{DARPA TIMIT acoustic-phonetic continous speech corpus CD-ROM.
  NIST speech disc 1-1.1},'' 1993.

\bibitem{barker2015chime}
J.~Barker, R.~Marxer, E.~Vincent, and S.~Watanabe, ``The third chime speech
  separation and recognition challenge: Dataset, task and baselines,'' in {\em
  IEEE Workshop on Automatic Speech Recognition and Understanding (ASRU)},
  pp.~504--511, 2015.

\bibitem{kingma2015adam}
D.~P. Kingma and J.~Ba, ``Adam: {A} {Method} for {Stochastic} {Optimization},''
  in {\em Int. Conf. on Learning Representations (ICLR)}, 2015.

\bibitem{pasad2021layerwise}
A.~Pasad, J.-C. Chou, and K.~Livescu, ``Layer-wise analysis of a
  self-supervised speech representation model,'' in {\em IEEE Workshop on
  Automatic Speech Recognition and Understanding (ASRU)}, pp.~914--921, 2021.

\end{thebibliography}

\end{document}